\documentclass[preprint,showpacs,preprintnumbers,amsmath,amssymb]{revtex4}

\usepackage{graphicx}
\usepackage{dcolumn}
\usepackage{bm}

\begin{document}

\preprint{}

\title{Larkin-Ovchinnikov state in resonant Fermi gas}
\author{Nobukatsu Yoshida and S.-K. Yip}
\affiliation{ Institute of Physics, Academia
Sinica, Nankang, Taipei 115, Taiwan}%

\date{\today}

\begin{abstract}

We construct the phase diagram of a homogeneous two component
Fermi gas with population imbalance under a Feshbach resonance.
In particular, we study the physics and stability of the
Larkin-Ovchinnikov phase.  We show that this phase is stable
over a much larger parameter range than what has been previously
reported by other authors.

\end{abstract}

\pacs{03.75.Ss, 05.30.Fk, 34.90.+q}
\maketitle

Trapped fermi gases offer us a wonderful opportunity to study strongly
interacting fermion systems \cite{expr}.  With Feshbach resonance, one is
able to tune the effective interaction from weakly attractive
for magnetic field above the resonance to strongly attractive below.
This problem is also closely related to studies of quark and nuclear matter
\cite{other-review}

Recent attention has shifted to systems with unequal populations
\cite{UNpd,SR06,SR-rev,unexpr}. The latter problem is analogous to
the physics of a superconductor under the influence of an external
Zeeman field, which provides a chemical potential difference
$\mu_{\uparrow} - \mu_{\downarrow} \equiv 2h$ between the two
species $\uparrow$ and $\downarrow$.  In the weak-coupling limit, it
was shown by Sarma \cite{Sarma} that the uniform state with
population imbalance is unstable at zero temperature. By comparing
the free energy of the normal state and the completely paired
superconducting state, he concluded that, as the magnetic field is
increased, there is a first order phase transition from the
equal-population superconducting state (with gap $\Delta_0$) to the
normal state at $h = \Delta_0/ \sqrt{2}$. This implies that a system
with given unequal numbers of the two spin species will either phase
separate(with $h = \Delta_0/\sqrt{2}$), or be in the normal state
(with $h > \Delta_0/\sqrt{2}$), depending on the imbalance. However,
later, Fulde-Ferrell \cite{FF} and Larkin-Ovchinnikov \cite{LO}
showed that there are better alternatives. Fulde and Ferrell (FF)
considered a state with order parameter $\Delta (\vec r) = \Delta_0
e^{ i \vec q \cdot \vec r}$, that is, pairs with finite momentum.
The order parameter has a spatially varying phase with however the
magnitude still constant in space. They showed that in certain
parameter space this state has lower free energy than the states
considered by Sarma.    Larkin and Ovchinnikov (LO), however
demonstrated that, at least in the small order parameter limit,
states with certain choices of sinusoidal variations of order
parameter (such as $\Delta (\vec r) = \Delta_0 {\rm cos} (q x)$ etc)
are more energetically favorable than the FF state.  Notice that in
the LO state, the magnitude of the order parameter is no longer a
constant in space. With decreasing magnetic field and hence
decreasing population difference $n_d$, many authors showed that the
LO state evolves into a state with a set of domain walls. This has
been demonstrated for one \cite{1D}, two \cite{Burkhardt}, as well
as in three \cite{Nagai} dimensions and also for a d-wave
superconductor \cite{Vorontsov}. In the small population imbalance
limit, the order parameter has a constant magnitude almost
everywhere in space (with value identical to the state with no
population difference), except near the domain walls. The phase of
the order parameter changes by $\pi$ when these walls are crossed,
and these are also the locations where the magnetization
concentrates. This local magnetization arises from the occupation of
bound states, available due to the presence of the domain walls. The
physics of these domain walls is closely related to the
$\pi$-junctions in SFS (S=superconductor, F=ferromagnet) junctions
\cite{Buzdin}, where for suitable parameters it is energetically
favorable for the two superconductors to acquire a phase difference
$\pi$.

Indeed, in three-dimensions, in the zero temperature and weak-coupling limit,
Matsuo et al \cite{Nagai} demonstrated that the energy of the domain walls
becomes negative for $h$ above $h_{dw} \equiv \frac{1}{2} \times
  0.745 \pi T_c \approx 0.665 \Delta_0$ where $\Delta_0$ is the
zero temperature gap for the completely-paired superfluid
of equal populations.  This $h_{dw}$ is less than
the phase separation field $h_{ps} = \Delta_0 / \sqrt{2} \approx 0.707 \Delta_0$
(where the free energy of the normal and the completely-paired
superfluid states are equal).
Hence the LO state must be more stable
than both the uniform superfluid and the normal state
(at least) for $h_{dw} < h < h_{ps}$.  For
$h$ slightly above $h_{dw}$, the domain walls are far apart.  The system
resembles the uniform paired state except for the occasional domain walls.
This state thus allows for the possibility of arbitrarily small population
difference between the two species.  This problem is quite analogous
to the lower critical field $h_{c1}$ in type II superconductors,
where the vortex energy becomes negative and vortices begin to penetrate
the superconductor for $h$ slightly above $h_{c1}$.
  For increasing $h$, more and more domain walls are formed,
the order parameter becomes more sinusoidal like and the state evolves
smoothly to the picture given by LO \cite{LO}.

For the resonant Fermi gas, it has been recognized that at
intermediate coupling strengths, the uniform state with population
imbalance must be unstable \cite{UNpd,SR06}.  Indeed this has also
been demonstrated also by experiments \cite{unexpr}.   However, the
investigation into the actual phase diagram, that is, the question
as to which phase appears where instead of the unstable uniform
state, even for the case without a trap, cannot be regarded as
complete, especially regarding the stability of the FFLO states.
Many papers \cite{Parish07,Gu06} examined only phase separation,
whereas some \cite{SR06,SR-rev,Hu06,He06,YanHe07} considered in
addition only the FF state with $\Delta \propto e^{i \vec q \cdot
\vec r}$.  However, none of these works actually investigated the LO
states. They conclude that the FF state exists only in a very narrow
region next to the normal state in the weak-coupling regime, and
deduce then that phase-separation occurs throughout the rest of the
entire region where the uniform phase is unstable. In particular,
they conclude that, for small population differences, phase
separation occurs unless the dimensionless coupling constant $1/k_f
a \to - \infty$. However, as seen already in the above paragraph,
this is simply an artifact of the FF state, which cannot smoothly go
into the completely paired equal population state \cite{Takada}. We
expect that at least, in the weak-coupling limit, the state should
be LO for any population imbalance below that of the normal state.

In this work, we shall investigate the stability of the LO state for
arbitrary strength of the attractive interaction between the two
species, thus generalizing previous works such as
\cite{Burkhardt,Nagai} beyond the BCS limit.  We shall concentrate
in particular on the small $n_d$ limit.  More specifically, we
compare the critical chemical potential difference $h_{dw}$ at which
the domain wall energy becomes negative, to the critical field
$h_{ps}$ for phase separation where the free energy of the normal
phase becomes equal to that of the completely paired equal
population superfluid phase.   We find that for $1/k_f a \lesssim
(\gtrsim) -0.845 $, $h_{dw} < (>) h_{ps}$. Therefore we conclude
that, for small $n_d$, the LO state is more stable than the phase
separated state for $1/k_f a \lesssim -0.845  $.  By combining with
previous results \cite{UNpd,SR-rev,UN-reply}
 (and with some reasonable extrapolations),
we sketch the appropriate phase diagram for our system.

 The mean-field Hamiltonian of our system can be written as

\begin{equation}
H = \int d^3{\vec r} \left\{
    \sum_{\sigma} \left[ \frac{ \hbar^2 \nabla \psi_{\sigma}^{\dagger}
  \nabla \psi_{\sigma}} { 2m}
   - \mu_{\sigma} \psi_{\sigma}^{\dagger} \psi_{\sigma} \right]
  - \left[ \Delta^{*} (\vec r) \psi_{\downarrow} \psi_{\uparrow} + c.c. \right]
  - \frac{|\Delta(\vec r)|^2} {g}
   \right\}
\label{ham}
\end{equation}
where $\sigma = \uparrow$, $\downarrow$ for the two species,
$\psi_{\sigma}(\vec r)$ their corresponding field operators,
$\Delta (\vec r)$ a position dependent
order parameter and $g$ the coupling constant.  For convenience below
we shall also write $\mu_{\uparrow} = \mu+h$ and $\mu_{\downarrow} = \mu-h$.

This Hamiltonian can be diagonalized by the Bogoliubov transformation for
a general inhomogeneous system \cite{deGennes}:
\begin{eqnarray}
\left( \begin{array}{c}
\psi_{\uparrow} (\vec r) \\
\psi_{\downarrow}^{\dagger} (\vec r)
\end{array} \right) =
  \sum_J
\left( \begin{array}{cc}
u_J(\vec r) & v^{*}_J (\vec r) \\
-v_J(\vec r) & u_J^{*} (\vec r)
\end{array} \right)
\left( \begin{array}{c}
\alpha_J \\
\beta_J^{\dagger}
\end{array} \right)
\label{transf}
\end{eqnarray}
where $\alpha_J$, $\beta_J$ are annihilation operators for
quasiparticles with spin $\uparrow$ and $\downarrow$
of the state labelled by a set of quantum numbers $J$.
$u_J (\vec r)$, $v_J (\vec r)$ obeys
$\int d^3{\vec r} \left[ |u_J(\vec r)|^2 + |v_J(\vec r)|^2 \right] = 1$
and the Bogoliubov-deGennes (B-dG) equation
\begin{eqnarray}
\left( \begin{array}{cc}
-\frac{\hbar^2 \nabla^2}{2m} - \mu & \Delta(\vec r) \\
\Delta^*(\vec r) & \frac{\hbar^2 \nabla^2}{2m} + \mu
\end{array} \right)
\left( \begin{array}{c}
u_J(\vec r) \\
v_J(\vec r)
\end{array} \right)
=
E_J
\left( \begin{array}{c}
u_J(\vec r) \\
v_J(\vec r)
\end{array} \right)
\label{bdG1}
\end{eqnarray}

Substituting eq (\ref{transf}) into eq (\ref{ham})
and using eq (\ref{bdG1}) to eliminate the derivatives
$\nabla u_J$ and $\nabla v_J$'s, we obtain the
ground state free energy
\begin{equation}
F = \int d^3{\vec r} \left\{
\sum_J \left[ - 2 E_J |v_J(\vec r)|^2 +
\frac{1}{V} \frac{|\Delta(\vec r)|^2}{2 \epsilon_J} \right]
  - \frac{m}{4 \pi \hbar^2 a} |\Delta(\vec r)|^2 \right\}
 + \sum_J (E_J - h) f (E_J -h)
\label{free1}
\end{equation}
The last term arises from the occupation of states $\alpha_J$ with
$E_J < h$.
Here $f$ is the Fermi function, $V$ the volume, and we have
eliminated the interaction constant $g$ in favor of
the scattering length $a$ via the relation
$\frac{1}{g} = \frac{m}{(4 \pi \hbar^2 a)}
 - \frac{1}{V} \sum_J \frac{1}{2 \epsilon_J}$
where $\epsilon_J$ is the energy of the state $J$ in the normal phase.
(Strictly speaking we should label the quantum states in the normal phase
by another set of quantum numbers $J'$, but we shall not make such
a distinction for simplicity in notations.  See further below.)
In eq (\ref{free1}), $\Delta(\vec r)$ should be viewed as a variational
parameter with respect to which $F$ has to be minimized.

  We are interested in the domain wall energy for given $\mu$, $h$ and $a$.
We evaluate this by calculating the energy difference between a
state with a single planar domain wall and the uniform completely
paired superfluid state. The latter is easy, since it is independent
of $h$ and the B-dG equation (\ref{bdG1}) can be solved by Fourier
transform. We obtained, for wavevector $\vec k$, the familiar
quasiparticle energies $E_{\vec k} = \left[ (\epsilon_k - \mu)^2 +
|\Delta|^2 \right]^{1/2}$ (where $\epsilon_k \equiv \frac{\hbar^2
k^2} { 2 m}$) and hence the bulk free energy $F_S (\mu) =
\sum_{\vec k} [ \epsilon_k - \mu - E_k
 + \frac{m}{\hbar^2 k^2} |\Delta|^2  ]
 - \frac{m}{4 \pi \hbar^2 a} |\Delta|^2$.
Minimizing with respect to $\Delta$ gives the usual BCS gap equation
$- \frac{  m }{ 4 \pi \hbar^2 a} \Delta \, =\, \Delta \frac{1}{V}
\sum_{\vec k} \left [ \frac{ 1 }{ 2 E_k } - \frac{  m }{ \hbar^2
k^2} \right ]$. We can also compute the corresponding density via $n
= \int \frac{d^3 k}{ (2\pi)^3}
  \left[
   1 - \frac{\epsilon_k  - \mu}{E_k} \right] $ and
the corresponding Fermi wavevector $k_f \equiv ( 3 \pi^2 n)^{1/3}$
and express $a$ in the dimensionless combination $k_f a$.
These results are identical with those in \cite{Engelbrecht}.
We had also used $\Delta(z) = {\rm const}$ instead of a domain
wall in the procedure described below and verified numerically that
we indeed obtained the same free energy density for the uniform state.
[Equating this free energy to that of the normal state at the same
$\mu$ and $h$, we obtained the phase separation field $h_{ps}$ discussed.]

  Next we evaluate the free energy of a planar domain wall.  For this,
we put our system in a box of dimensions $L_x$, $L_y=L_x$, $L_z \equiv L$.
We assume that the order parameter varies only along the $z$ direction
and thus Fourier transform the $x$ and $y$ coordinates.
Calling the resulting wavevector $\vec p$,
we rewrite $(u_J,v_J)$ as
\begin{eqnarray}
\left( \begin{array}{c}
u_J(\vec r) \\
v_J(\vec r)
\end{array} \right)
=
\frac{1}{L_x L^{1/2} }
\left( \begin{array}{c}
u_{p,j}(z) \\
v_{p,j}(z)
\end{array} \right)
e^{i \vec p \cdot \vec r_p}
\end{eqnarray}
where $\vec r_p$ is the component of $\vec r$ in the x-y plane,
and $j$ is now a quantum number for the $z$ dependences.
$u_{p,j}(z), v_{p,j}(z)$ are dimensionless quantities obeying
$\int dz \left( |u_{p,j}(z)|^2 + |v_{p,j}(z)|^2 \right)/L = 1$.
Eq (\ref{bdG1})
becomes
\begin{eqnarray}
\left( \begin{array}{cc}
-\frac{\hbar^2 \partial_z^2}{2m} - \tilde \mu & \Delta(z) \\
\Delta^*(z) & \frac{\hbar^2 \partial_z^2}{2m} + \tilde \mu
\end{array} \right)
\left( \begin{array}{c}
u_{p, j} (z) \\
v_{p, j}  (z)
\end{array} \right)
=
E_{p,j}
\left( \begin{array}{c}
u_{p,j}(z) \\
v_{p,j}(z)
\end{array} \right)
\label{bdG2}
\end{eqnarray}
where $\tilde \mu \equiv \mu - \frac{p^2}{2m}$.
The energies $E_{p,j}$ and the wavefunctions
$u_{p,j}(z)$, $v_{p,j}(z)$ depend on $\vec p$ only
through its magnitude $p$.
The free energy can then be written as
\begin{equation}
F = \int dz \left\{ \frac{1}{L}
\sum_{\vec p} \sum_j^{j_{max}} \left[ - 2 E_{p,j} |v_{p,j} (z)|^2 +
\frac{|\Delta(z)|^2}{2 \epsilon_{p,j}} \right]
  - L_{x}^2 \frac{m}{4 \pi \hbar^2 a} |\Delta(z)|^2 \right\}
 + \sum_{\vec p,j} (E_{p,j} - h) f (E_{p,j} -h)
\label{free2}
\end{equation}

In this equation, we arrange the quasiparticle states for both
the superfluid and the normal states as increasing function of the
counting number $j$.
We solve eq (\ref{bdG2}) by discretizing the $z$ coordinate.
After the energies and the function $v_{p,j} (z)$ are calculated,
we put it in eq (\ref{free2}) and calculate the free energy.
For $\Delta(z)$,
we limit ourselves to the one-parameter ansatz
\begin{equation}
\Delta(z) = \Delta_0 {\rm tanh} \left( \frac{k_{\mu} z}{\beta} \right)
\label{ansatz}
\end{equation}
with $\beta$ as the variational parameter and $k_{\mu} \equiv ( 2 m \mu)^{1/2}$.
(Thus the width of the domain wall is given by $\beta k_{\mu}^{-1}$.)
Here $\Delta_0$ is
the bulk order parameter for the given $\mu$ and $a$ at $h=0$,
as the order parameter should approach this value far away
from the domain wall.  Our eq (\ref{ansatz}) is motivated
by earlier investigations \cite{Burkhardt,Nagai}, where their
numerical results can be well fitted by a function of the form (\ref{ansatz}).
(Since we are employing periodic boundary conditions in $z$, this ansatz
actually introduces a sharp ($\beta=0$) domain wall also at $z=\pm L/2$.
However, this can be taken care of easily by removing the
contributions due to this extra domain wall.)

We shall then input the ansatz eq (\ref{ansatz}) into eq (\ref{bdG2}) to
solve for $E_{p,j}$.  Our analysis of the free energy
eq (\ref{free2}) is simplified
by the following observations.  At $h=0$, since all quasiparticle
energies are positive, $f(E_{p,j}) = 0$ and the free energy is
given simply by the integral in eq (\ref{free2}).  The free energy at
finite $h$ of a given $\beta$ is related to that of $h=0$ at the same $\beta$
by simply adding the negative term $\sum_{\vec p,j} (E_{p,j} -h) f(E_{p,j} -h)$
due to the occupation of the quasiparticle states.

\begin{figure}[tbh]
\begin{center}
\includegraphics[width=4in]{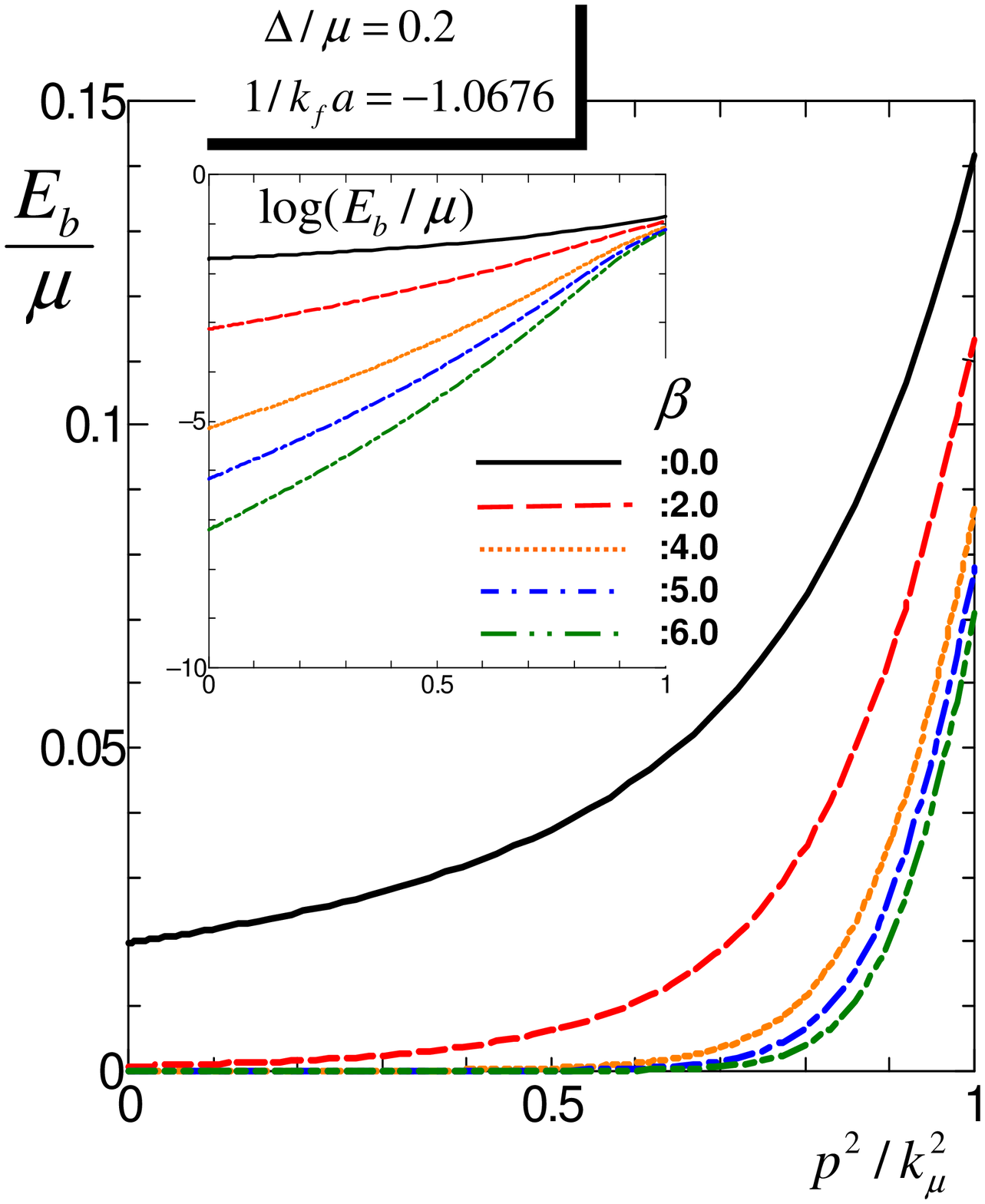}
\end{center}
 \caption{(color online) Bound state energies for
$1/k_f a = -1. 0676$, corresponding to
$1/k_{\mu} a = -1.0866$, $\Delta_0/\mu =0.2$,
for widths of the domain walls given by $\beta k_{\mu}^{-1}$.
Energies are normalized to $\mu$ and the momentum p in
the x-y plane normalized to $k_{\mu}$.  Inset:
  bound state energies in log plot.}
 \label{fig:bound}
 \vspace{-5pt}
\end{figure}

An example for our results for $1/k_{\mu} a = -1.0866$, 
$1/k_f a = -1.0676$
and $\Delta_0/\mu = 0.2$ are discussed below.
The bound state energies $E_b \equiv E_{p,j}$
are illustrated in Fig \ref{fig:bound}.
We see that in general we have states below the continuum
($E_{p,j} < |\Delta_0| $ for $\tilde \mu > 0$,
 $E_{p,j} < \sqrt{|\tilde \mu|^2 + |\Delta_0|^2} $ for $\tilde \mu < 0$.
It is the bound states that are essential:  as we shall see,
the relevant values of $h$ are below the gap edges).
The
$\beta=0$ results were checked against the analytical ones in the Appendix.
The bound state energy for a given $p$ decreases with the width of the domain walls,
as one expects.  The significance of this would be discussed again below.

\begin{figure}[tbh]
\begin{center}
\includegraphics[width=4in]{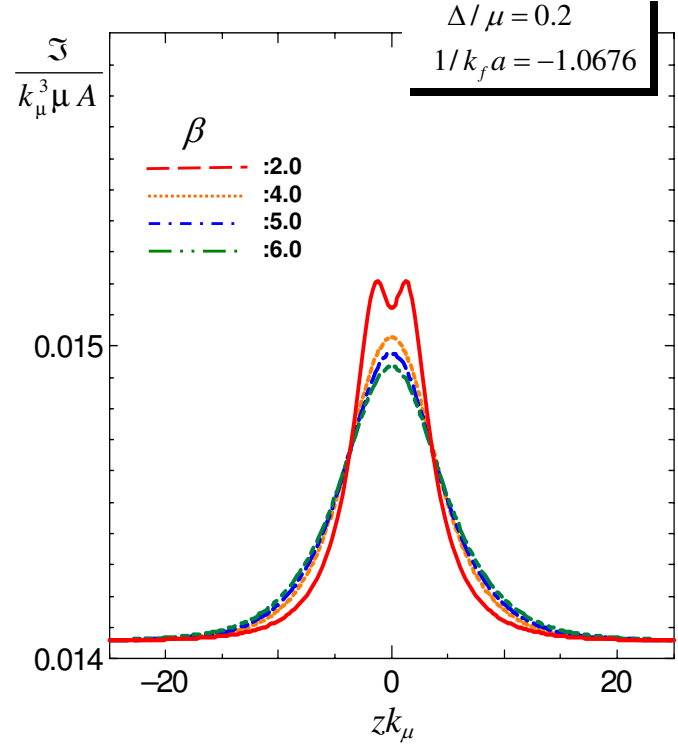}
\end{center}
\vspace{-5pt}
 \caption{(color online) Integrand of eq (\ref{free2}),
in units of $k_{\mu}^3 \mu A$ where $A$ is the area,
 as a function of position $z$ (in units of $k_{\mu}^{-1}$).
  Parameters are the same as in Fig \ref{fig:bound}.}
 \label{fig:fz}
 \vspace{-5pt}
\end{figure}

\begin{figure}[tbh]
\begin{center}
\includegraphics[width=4in]{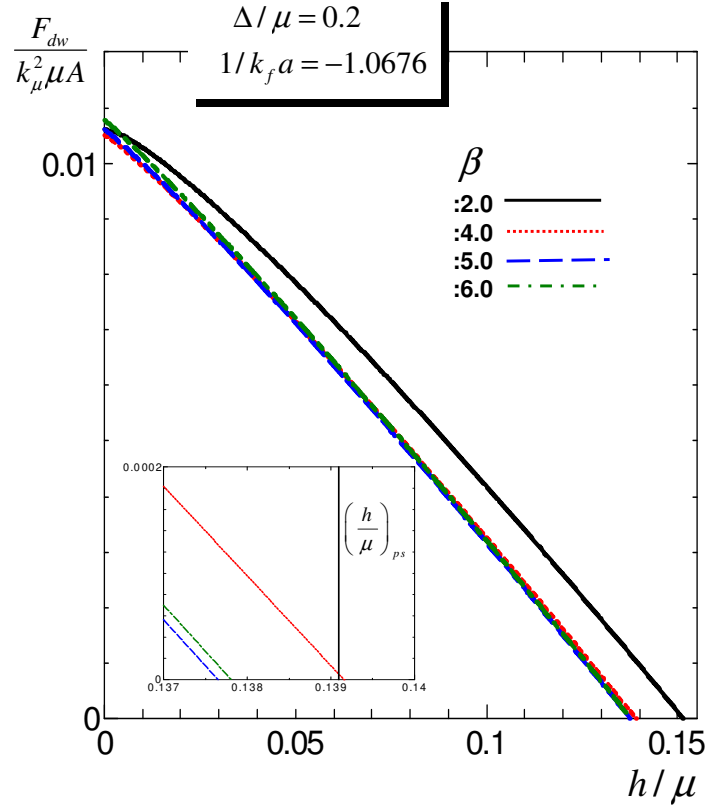}
\end{center}
 \caption{(color online) Domain  wall free energy $F_{dw}$ per unit area $A$
(in units of $k_{\mu}^2 \mu$)
versus $h$ (in units of $\mu$).
  Parameters are the same as in Fig \ref{fig:bound}.
Inset magnifies the region near $F_{dw} = 0$. }
 \label{fig:free-h}
 \vspace{-5pt}
\end{figure}

\begin{figure}[tbh]
\begin{center}
\includegraphics[width=4in]{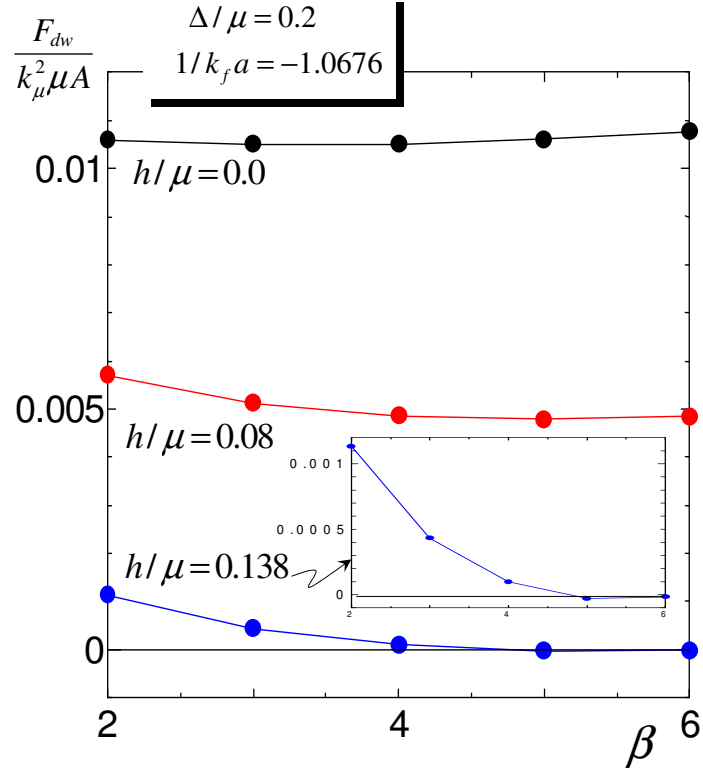}
\end{center}
 \caption{(color online) Domain wall free energy $F_{dw}$ per unit area $A$
(in units of $k_{\mu}^2 \mu$)
versus $\beta$.
  Parameters are the same as in Fig \ref{fig:bound}.
Inset shows the detailed behavior for $h/\mu = 0.138$.}
 \label{fig:free-b}
 \vspace{-5pt}
\end{figure}

The integrand of the first term in eq (\ref{free2}) is plotted in
Fig \ref{fig:fz}. (The double hump structure for small $\beta$ ($\ne
0$) is due to the fact that $|\Delta(z)| \to 0$ as $z \to 0$: see eq
(\ref{free2})). As expected, the integrand decreases to a constant
corresponding to the free energy density in the bulk for distances
sufficiently far away from the domain wall.  We can then evaluate
the domain wall energy $F_{dw}$ at $h=0$ by simply integrating this
excess contribution over $z$.
  At $h=0$, $F_{dw}$ is minimum at $\beta \approx 3$.
$F_{dw}$ is positive for all $\beta$'s, as expected since the
uniform state should have a lower energy than a domain wall. For
finite $h$, the free energies are evaluated by adding the negative
term from bound state occupation as discussed above. The results are
depicted in Fig \ref{fig:free-h}. The free energy decreases due to
the occupation of the bound states. Since the bound state energies
are smaller for larger $\beta$, the wall energy decreases faster for
larger $\beta$, shifting the $\beta$ for minimum wall energy to
larger values with increasing $h$ (see Fig \ref{fig:free-b}).
 The domain wall energy becomes
negative at sufficiently large $h$ for all $\beta$'s.   The important question is
whether it will become negative for some $\beta$ at a value of $h$ which
is less than that for phase separation.  For the parameters under discussion,
the domain wall energy first becomes negative for $\beta \approx 5$ at
$h \equiv h_{dw} \approx 0.1376 \mu$, which is less than the phase
separation value $h_{ps} \approx 0.1391 \mu$.  We thus conclude that
for $1/k_f a = -1.0676$, when $n_d$ is infinitesimally small, the
system is in the LO state but not the phase separated state.

At this point, it is of interest to compare our results more quantitatively
with those in \cite{Nagai}, who has assumed the quasiclassical limit
in their calculations at the outset.  They obtained the critical value
for the sign change of the domain wall energy at $h = h_{dw}
 = \frac{1}{2} \times 0.745 \pi T_c \approx 0.665 \Delta_0$
(using $\Delta_0 = 1.76 T_c$).  Our $h_{dw}$ is close to their value
if we simply rewrite their result as $h_{dw}/\mu = 0.665 \Delta_0/\mu$ and
substitute $\Delta_0/\mu = 0.2$.  Also, their Fig 1 indicates a domain wall
width near the critical field to be
roughly given by $ 2 \Delta_0 v_f / ( \pi T_c)^2$
where $v_f = k_f/m$ is the Fermi velocity.
Rewriting this expression as
$ \approx 1.25 \left( \frac{k_f}{k_\mu} \right)
 \left( \frac{\mu}{\Delta_0} \right) k_{\mu}^{-1}$
and simply subsituting the values of $\Delta_0/\mu$ etc we again find very
good agreement.  Thus, even though we did not assume the quasiclassical
limit at the outset, for $1/k_f a = -1.067$ our results can be understood
from the quasiclassical limit calculations of \cite{Nagai} with extrapolations.

\begin{figure}[tbh]
\begin{center}
\includegraphics[width=4in]{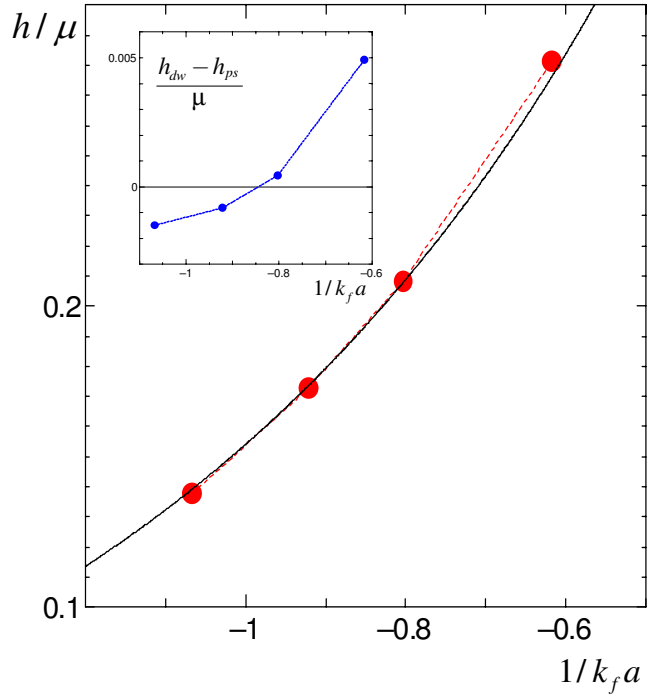}
\end{center}
 \caption{(color online)
  Comparison between $h_{dw}$ (dotted line)
and $h_{ps}$ (full line) as a function of
$1/k_f a$.  Inset: $h_{dw} - h_{ps}$. }
 \label{fig:h-ka}
 \vspace{-5pt}
\end{figure}

\begin{figure}[tbh]
\begin{center}
\includegraphics[width=5in]{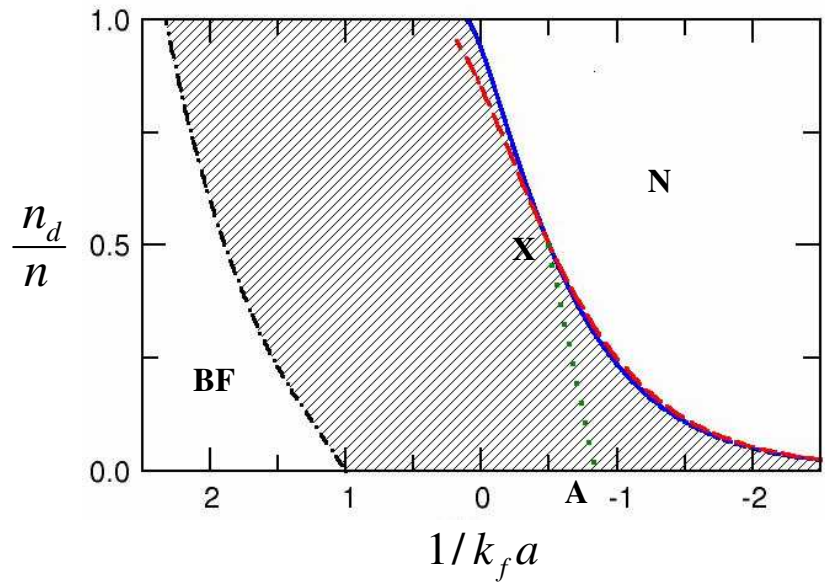}
\end{center}
 \caption{(color online)
  Phase diagram.  N: normal state, BF: homogeneous superfluid
(Bose-Fermi mixture) phase.
 Shaded regions to the right of XA: Larkin-Ovchinnikov state,
 shaded regions to the left of XA: phase separation.}
 \label{fig:pd}
 \vspace{-5pt}
\end{figure}

For increasing $1/k_f a$, both $h_{dw}$ and $h_{ps}$ increase (and
the $\beta$ for the optimal domain wall decreases).
  However, $h_{dw}$ increases faster than $h_{ps}$,
and eventually $h_{dw} < h_{ps}$ no longer holds.
The situation is shown in
Fig \ref{fig:h-ka}.  If $h_{dw} > h_{ps}$, then
the formation of domain wall is no longer favorable
since phase separation already occurs at $h$
slightly above $ h_{ps}$.
We conclude then, for $n_d \to 0_{+}$, the transition
from the LO state to the phase separation state occurs
at $1/k_f a \approx -0.845$.  This point is indicated by
the point A in Fig \ref{fig:pd}.

After locating this transition point for $n_d \to 0_{+}$,
 we now attempt to construct
a phase diagram for general $n_d$ by combining the present results
with previous ones in the literature.
For $n_d/n \ne 0$ and sufficiently negative $1/k_f a$, the system is
in the normal state \cite{UNpd,SR-rev}.  We consider in turn the
transition lines between this normal state and the LO and phase
separated states. Assuming that the transition to the LO state is
second order \cite{LO} (c.f. below), the transition line between the
normal and LO state can be found by solving the Cooper problem at
finite wavevector $\vec q$:

\begin{equation}
- \frac{  m }{ 4 \pi \hbar^2 a} \, =\,
\frac{1}{V} \sum_{\vec k}
\left [ \frac{ 1 - f(\epsilon_{\vec k + \vec q} - (\mu + h))
  - f(\epsilon_{- \vec k} - (\mu - h)) }
{ \epsilon_{\vec k + \vec q} + \epsilon_{\vec k} - 2 \mu }
- \frac{  m }{ \hbar^2 k^2} \right ]
\label{Cq}
\end{equation}

The transition line is determined by finding the optimal $q$
corresponding to the weakest attractive interaction,
{\it i.e.}, the most negative $1/k_f a$.
We note here that, since eq (\ref{Cq}) is obtained by
linearizing in the order parameter, the same equation
determines the second order transition line into the FF state
\cite{LO}.  Our result can be turned into a line in our
phase diagram (an equation
of $1/k_f a$ versus $n_d/n$) by using
$k_f = ( 3 \pi^2 n)^{1/3}$, with
$n = \frac{ ( 2m)^{3/2}}{ 6 \pi^2}
 \left[ (\mu+h)^{3/2} + (\mu-h)^{3/2} \right]$
and $\frac{n_d}{n} =
\frac{ (\mu+h)^{3/2} - (\mu-h)^{3/2}}
 {(\mu+h)^{3/2} + (\mu-h)^{3/2} } $.
This gives the long-dashed line in Fig \ref{fig:pd}.
Our numerical results (see also \cite{UN-reply}) agree well with \cite{SR-rev}.
It should however be remarked that the transition from
the normal state to the LO state has also claimed to
be first order \cite{Mora05} at very low temperatures
in three-dimensions. However, the difference
between the actual transition line and
the second order line  is very small \cite{Mora05},
and we ignore this difference here.

For the transition line between the normal state and phase
separation, we equate \cite{Sarma} the free energy of the completely
paired superfluid state $F_S (\mu)$ to that of the normal state:
$F_N( \mu, h) = - V \frac{ (2m)^{3/2}}{15 \pi^2}
 \left[ (\mu+h)^{5/2} + (\mu-h)^{5/2} \right]$.
This again yields a line $1/k_f a$ as a function of $h/\mu$
and hence $n_d/n$, shown as the solid line in fig \ref{fig:pd}.
Our numerical result (see also \cite{UN-reply}) is also in good agreement with
\cite{SR-rev}.  The two transition lines intersect at the point
labelled by X
at $1/k_f a \approx -0.50$.  Therefore the transition
is to the LO (phase separation) state if
$1/k_f a$ is less (larger) than this value.
Interpolating between points A and X, we thus conclude
that the system is in the LO state for the shaded region
to the right of the line XA, whereas phase separation
occurs for the shaded region to the left of this line.
The transition lines between this phase-separated state
and the homogeneous superfluid (Bose-Fermi mixture) phase
for positive $1/k_f a$ have already been discussed
elsewhere in the literature ( \cite{SR-rev,Parish07,Gu06},
see also other references in \cite{UN-reply}),
and we would not repeat the details here.

In conclusion, we studied the stability of the LO state in a
homogeneous system, in particular in the limit of small population
imbalance, by calculating the domain wall energies. Our
investigation here is analogous to the determination of the lower
critical field of the vortex phase of a type II superconductor.  The
determination of the detailed structures of the LO state, such as
the question of the lattice structure of the domain walls (c.f. e.g.
\cite{Mora05}, analogous to the vortex lattice structure in type II
superconductors), as well as the phase diagram in the presence of a
trap (studied already partially in \cite{LOtrap}), are left for the
future.

We thank C.-H. Cheng, C.-H. Pao and S.-T. Wu for
their help in obtaining the transition lines from
the normal state.
This research was supported by the National Science
Council of Taiwan under grant number
 NSC95-2112-M-054-MY3.

\vskip 1 cm

{\bf Appendix -- Bound states}

Due to the important role played by the bound states
in the LO state, we give the analytic results for  a sharp domain wall
where $\Delta (z) = \Delta_0 sgn(z)$ (corresponding to $\beta \to 0$).
It should be noted that, even though $|\Delta(z)| $ is a constant
throughout in space, bound states still exist due to the sudden
change in phase factor of the order parameter from $-\pi$ to $0$ at $z=0$.
  As already seen from eq (\ref{bdG2}),
the bound state energy is a function of $\mu$ and $p$ only through
the combination $\tilde \mu$. The B-dG equation (\ref{bdG2}) can be
solved easily in this case since $\Delta(z)$ is piecewise constant.
We require that the functions $u_{p,j}(z)$, $v_{p,j}(z)$ as well as
their derivatives be continuous at $z=0$.  The calculation is
similar to solving the Schr\"odinger equation in piecewise constant
potentials.   It is straightforward and we simply state the results.
It is convenient to divide them into two regions:

(i) $\tilde \mu > 0$.  In this case states are bound
when $E_b \equiv E_{p,j}  < |\Delta_0|$.  We find that there is only one bound state
(thus correspondingly one bound state for each $\vec p$), with energy given by
\begin{equation}
\frac{E_b}{|\Delta_0|} =   \frac{\sqrt{1 + 2 \delta^2} -1 } { 2 \delta}
\label{b1}
\end{equation}
where $\delta \equiv |\Delta_0|/\tilde \mu$.
We note here that the quasiclassical limit corresponds to
$|\Delta_0| \ll \tilde \mu$
hence $\delta \ll 1 $.
In this limit, one can approximate
the B-dG equation by the Andreev equation
\cite{Andreev64} and
the bound state energy vanishes
(irrespective of the detailed positional dependence
of $\Delta(z)$ so long as there is a
$\pi$ phase change: see e.g \cite{Lofwander} and also \cite{Atiyah}).
For small $|\Delta_0|/\tilde \mu$, we have
$E_b \approx |\Delta_0|^2/ 2 \tilde \mu$. (c.f. the
mini-gap for bound states near the
vortex core \cite{deGennes}).  With increasing $|\Delta_0|$ or equivalently
decreasing $\tilde \mu$, $E_b /|\Delta_0|$ increases.  For $\delta \to \infty$,
$E_b \to |\Delta_0|/ \sqrt{2}$.

(ii) $\tilde \mu < 0$.  In this case the continnuum starts at
$ \sqrt{ \tilde \mu^2 + |\Delta_0|^2 }$. Again we find only
one bound state, with
\begin{equation}
\frac{E_b}{|\Delta_0|} = \frac{
  \sqrt{ 1 + 2 \delta^2 } + 1 }{ 2 \delta }
\label{b2}
\end{equation}
where $\delta \equiv |\Delta_0|/|\tilde \mu|$. For $\tilde \mu =
0_{-}$ ($\delta \to \infty$), $E_b \to |\Delta_0|/ \sqrt{2}$ (c.f.
case (i) above). For decreasing $\tilde \mu$ (decreasing $\delta$),
$E_b$ increases. $E_b$ reaches $|\Delta_0|$ at $\delta = 2$. For
$\tilde \mu \to - \infty$, $E_b$ approaches the continuum from
below. It is remarkable that bound state exists even for $\tilde \mu
< 0$. We note that in this regime we have always $E_b > |\tilde
\mu|$. A similar situation occurs also for the bound states at a
vortex \cite{Sensarma} for general coupling strength $1/k_f a$.



\vspace{15pt}

\vspace{15pt}

\vspace{15pt}

\vspace{15pt}

\vspace{15pt}

\vspace{15pt}

\end{document}